\documentclass[apj]{emulateapj}%
\usepackage{apjfonts}
\usepackage{amsfonts}
\usepackage{amsmath}
\usepackage{amssymb}
\usepackage{graphicx}
\usepackage{comment}
\usepackage{natbib}%
\providecommand{\U}[1]{\protect\rule{.1in}{.1in}}
\newcommand{\pd}{\partial}
\begin{document}

\title{Theory of photospheric emission from relativistic outflows}
\author{R. Ruffini, I.~A. Siutsou and G.~V. Vereshchagin}

\begin{abstract}
Two popular models of optically thick relativistic outflows exist: the wind and
the shell. We propose a unified treatment of photospheric emission within these
models. We show that quite counterintuitive situations may appear when e.g.
geometrically thin shell may behave as thick wind. For this reason we introduce
notions of photon thick and photon thin outflows. They appear more general and
better physically motivated than winds and shells when photospheric emission is
considered.

We obtain light curves and observed spectra for both photon thick and photon
thin outflows. In the photon thick case we generalize the results obtained for
steady wind. It is our main finding that the photospheric emission from the
photon thin outflow is dominated by diffusion and produces non thermal time
integrated spectra, which may be described by the Band function well known in
the GRB literature.

Energetic GRBs should produce photon thin outflows and therefore when only time
integrated spectra for such GRBs are available we naturally expect them to have
a Band shape. In the literature Band spectra for the photospheric emission of
GRBs are obtained only involving additional dissipative mechanisms which are
not required here.
\end{abstract}


\shorttitle{Theory of photospheric emission from relativistic outflows}

\affil{ICRANet, 65122, p.le della Repubblica, 10, Pescara, Italy} \affil{ICRA
and Dipartimento di Fisica, Universit\`a di Roma ``Sapienza'', 00185, p-le A.
Moro 5, Rome, Italy}

\maketitle

\section{Introduction}

High Lorentz factors of the bulk motion of various outflows are common in
relativistic astrophysics. The best known examples are Active Galactic Nuclei
\citep{2003ASPC..290..275M}, microquasars and Gamma-Ray Bursts (GRBs)
\citep{2004RvMP...76.1143P}. In the latter case outflows indeed reach
ultrarelativistic velocities.

Various models are suggested to explain the acceleration of outflows to
ultrarelativistic velocities. The electromagnetic model
\citep{2006NJPh....8..119L}\ assumes that the energy in the source of GRB is
converted into electromagnetic energy which is transported in the form of a
Poynting flux. We adopt here another popular idea that the energy release leads
to creation of an optically thick source which expands due to thermal
acceleration. This idea is the basis of both the fireball
(\citet{1999PhR...314..575P,2004RvMP...76.1143P} and references therein) and
the fireshell (\citet{2001ApJ...555L.113R,2009AIPC.1111..325R} and references
therein) models.

In pioneer works by \citet{1986ApJ...308L..47G} who considered an instant
explosion, and by \citet{1986ApJ...308L..43P} who discussed a gradual energy
release, a conclusion was reached that the electron-positron plasma is created
in the source of GRB. Assuming further that the plasma reaches thermal
equilibrium they focused on hydrodynamic expansion in such models and gave
photometric and spectroscopic predictions for GRBs. Later, baryonic loading of
fireballs was considered for explosions by \citet{1990ApJ...365L..55S} and for
winds by \citet{1990ApJ...363..218P}. \citet{1991ApJ...369..175A} considered
the appearance of the photosphere of the relativistic wind to a distant
observer showing that its shape is concave.

The interest in photospheric emission from relativistic winds has been revived
recently in papers by \citet{2002MNRAS.336.1271D}, \citet{2007ApJ...664L...1P},
\citet{2011ApJ...737...68B}, \citet{2011arXiv1103.0708R},
\citet{2011ApJ...732...49P}, and others. In the fireshell model, which assumes
an explosive energy release, the first potentially visible component of any
GRB, the Proper GRB \citep{2001A&A...368..377B,2001ApJ...555L.113R}, comes from
the thermal flash of radiation emitted when the outflow becomes transparent for
photons.

One of the main scopes of this paper is to show that the association of
``instantaneous energy release'' with ``thin shell'' (e.g.
\citet{1993MNRAS.263..861P} and \citet{1993ApJ...415..181M}) and ``continuous
energy release'' with ``thick wind'' (e.g. \citet{1986ApJ...308L..43P,
1990ApJ...363..218P}) generally adopted in the literature is incomplete with
respect to the photospheric emission. A unified treatment of ultrarelativistic
outflows originating both from instantaneous and from continuous energy release
is presented here. We also propose a new classification of such outflows which
is complementary to the traditional division on shells and winds. When
photospheric emission is considered this new classification provides some
additional physical insights. Based on these results we present a semi-analytic
treatment for observed flux of photospheric emission assuming isotropic thermal
distribution of scattered photons in the comoving frame. We compute both
instantaneous and time-integrated spectra. These results are then applied
within both the shell and the wind models of GRBs. Remarkably the
time-integrated spectrum of energetic GRBs is predicted to have a Band shape
\citep{1993ApJ...413..281B}.

The structure of the paper is as follows. In Section 2 we discuss, compare and
contrast an impulsive explosion and gradual energy release, giving rise
respectively to an ultrarelativistic shell and wind. In Section 3 we present
the computation of the optical depth and photospheric radius of the
relativistic outflow and discuss applicability of recovered asymptotics to
GRBs. In Section 4 we discuss radiation diffusion in an expanding relativistic
outflow. Observed light curves and spectra of photon thick and photon thin
outflows are computed in Sections 5 and 6, respectively. In Section 7 we
discuss main results obtained in this work and their implications for GRBs.
Conclusions follow. Details of derivation and application of the radiative
diffusion approximation are given in Appendix.

\section{Optically thick relativistic outflows:\ wind vs.\ explosion}

\label{WindExpl}

Relativistic outflows are generally classified as \emph{winds} or \emph{shells}
depending on how fast the energy in their sources is released. Consider energy
release in a source of linear dimension $R_{0}$. If the time scale of energy
release is short $\Delta t\simeq R_{0}/c$, an explosion occurs, which may be
characterized by the size $R_{0}$, total energy released $E_{0}$ and total
baryonic mass $M$. Relativistically expanding material forms a shell having
width approximately $R_0$. When the energy is released gradually, on a time
scale $\Delta t\gg R_{0}/c$, but the source luminosity $L$ exceeds the
Eddington limit, a wind is formed, which is characterized by its activity time
$\Delta t$, luminosity $L$ and mass ejection rate $\dot{M}$.

In both wind and shell models of cosmological GRBs the region of energy release
$R_0$ is clearly macroscopic, being of the order of the size of compact
astrophysical objects, about $R_0\sim 10^8$ cm. Considering the isotropic
energy involved $E_0\sim 10^{54}$ erg the temperature of plasma in this region
is typically in MeV range. The time scale of thermalization in such plasma is
of the order of $10^{-12}$ sec \citep{2007PhRvL..99l5003A, Aksenov2009}, being
much shorter than the dynamical time scale $R_0/c$. Since such dense plasma is
optically thick thermal equilibrium is established prior to
expansion in both shell and wind models. Opacity in such plasma is dominated by
Compton scattering. When the temperature in the source of the relativistic
outflow is large enough for electron-positron pair creation, $e^{+}e^{-}$ pairs
make an additional contribution to the opacity.

Whether the outflow becomes relativistic or not depends on the entropy in the
region where the energy is released. Both the wind and explosion cases can be
parameterized \citep{1990ApJ...365L..55S} by a dimensionless entropy parameter
$\eta$, equal to $E_0/Mc^{2}$ for shell model and $L/\dot{M}c^{2}$ for wind
model. When the baryonic loading $B=\eta^{-1}$ \citep{2000A&A...359..855R} is
sufficiently small the baryons will be accelerated to a relativistic velocity
$v$\ of bulk motion, attaining large Lorentz factors up to $\Gamma=\left[
1-\left(  v/c\right) ^{2}\right] ^{-1/2}\simeq\eta$, while in the opposite case
of large baryonic loading the outflow remains nonrelativistic with $v\simeq
c/\sqrt{2\eta}$.

In what follows we consider only ultrarelativistic spherically symmetric
outflows with $\Gamma\gg1$. In the simplest cases of wind or explosion in
vacuum, the dynamics of the outflow is divided into an acceleration phase and
a coasting phase \citep{1999PhR...314..575P} with respectively%
\begin{align}
    \Gamma &  \simeq\frac{r}{R_{0}}, & n_c &  \simeq n_{0}\left(  \frac{r}{R_{0}%
        }\right)^{-3}, & R_{0}  <r<\eta R_{0},\label{acceleration}\\
    \Gamma &  \simeq\eta=\mathrm{const}, & n_c &  \simeq \frac{n_{0}}{\eta}\left(  \frac{r}%
        {R_{0}}\right)  ^{-2}, & r>\eta R_{0},\label{coasting}%
\end{align}
where $n_c$ is the comoving number density of baryons in the
outflow\footnote{All quantities with subscript "c" are measured in
comoving reference frame, and all quantities without this subscript are
measured in laboratory reference frame.}. Notice that in the case of an impulsive
explosion for $r\gg R_{0}$ the matter and energy appear to a distant observer
to be concentrated in a \emph{geometrically thin} shell having width $l\sim
R_{0}$ due to the relativistic contraction \citep{1993MNRAS.263..861P}.

It is important to stress that both \emph{an infinitely long wind} with a
time-independent mass ejection rate and luminosity on the one hand, and
\emph{an infinitely thin shell} originating from an instantaneous explosion in
infinitely thin region represent two limiting cases for the energy release.

During both acceleration and coasting phases the continuity equation for the
laboratory number density reduces to $n\propto r^{-2}$. We take for the
laboratory density profile
\begin{equation}\label{nxi}
    n=\left\{
        \begin{array}[c]{cc}
            n_{0} \left( \dfrac{R_{0}}{r}\right) ^{2},
               & R(t)<r<R(t)+l,\\
            \  & \\
            0, & \mathrm{otherwise,}
        \end{array}
    \right.
\end{equation}
where $R(t)$ is the radial position of the inner boundary of the outflow.
Such an outflow may be produced by a gradual energy release with
constant luminosity and mass ejection rate on a finite time $\Delta t$ and we
will refer to it as the \emph{portion of wind}.

Below we show that both the shell and the wind defined above may appear for
photons emitted inside it as long wind or as thin shell, depending on the
initial conditions that specify respectively their width $l$ and activity
duration $\Delta t$. This appearance is a consequence of ultrarelativistic
expansion of the outflow. It is crucial to keep in mind that photons emitted
inside the expanding outflow propagate in a medium whose laboratory number
density depends both on radial coordinate and on time $n(r,t)$. For photons
propagating in the wind the spatial dependence of the number density plays the
key role, while for photons propagating in the shell its time dependence is
crucial.

\section{Optical depth and photospheric radius}
\label{optdepth}

The optical depth along the photon world line $\mathcal{L}$ is defined as
\begin{equation}\label{tauWL}
    \tau=\int_{\mathcal{L}}\sigma j_{\mu}dx^{\mu},
\end{equation}
where $\sigma$\ is Thomson cross section, $j^{\mu}$ is the 4-current of
particles, on which the photon scatters, and $dx^{\mu}$ is the element of the
photon world line.

\begin{figure}[ptb]%
\centering
\includegraphics[width=\columnwidth]{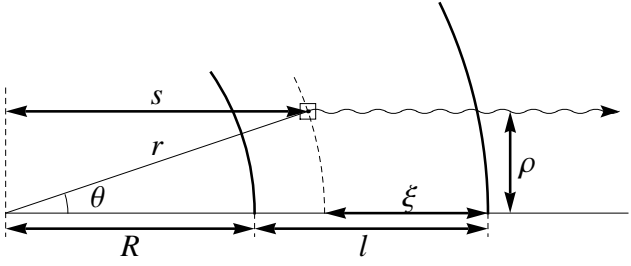}
\caption{Geometry of the outflow and variables used.
Observer is located to the right at infinity.
\label{xi}}
\end{figure}

Consider a spherically symmetric expanding outflow with an ultrarelativistic
velocity $v=\beta c\simeq1-{1}/{2\Gamma^{2}}$. Assume that the photon is
emitted at time $t$ from the interior boundary $r=R$ of the outflow and it
propagates outwards. This assumption is relaxed in Section \ref{geomdyn}. The
optical depth computed along the photon path from (\ref{tauWL}) is (see e.g.
\cite{1991ApJ...369..175A})
\begin{equation}
\tau=\int_{R}^{R+\Delta R}\!\!\!\sigma n\left( 1-\beta\cos\theta\right)
\frac{dr}{\cos\theta},\label{tau}%
\end{equation}
where $R+\Delta R$ is the radial coordinate at which the photon leaves the
outflow, and $\theta$ is the angle between the velocity vector of the outflow
and the direction of propagation of the photon, $n$ is the laboratory number
density of electrons and positrons, which may be present due to pair
production. Geometry of the outflow and used variables are illustrated by Fig.
\ref{xi}.

A pure electron-positron plasma reaches thermal equilibrium before expansion
\citep{2007PhRvL..99l5003A, Aksenov2009} and it remains accelerating
until it becomes transparent to radiation. Due to exponential dependence of
thermal pair density on the radial coordinate transparency is reached at
\begin{equation}\label{Ttrpair}
    kT_{\pm}\simeq0.040m_{e}c^{2}%
\end{equation}
rather independent of the initial conditions. Note that the optical depth for an
expanding electron-positron-photon shell computed by
\citet{1990ApJ...365L..55S} is incorrect since it uses photon thin
asymptotics, see 3rd line in (\ref{tau2}) below, which never applies to the pure
$e^{+}e^{-}$ outflows. The formula (\ref{Ttrpair}) is in agreement with works
of \cite{1998MNRAS.300.1158G} and \cite{2008ApJ...677..425L}.

Electron-positron plasma with baryonic loading also reaches thermal equilibrium
before its expansion starts \citep{2009PhRvD..79d3008A, Aksenov2009}. With
decreasing entropy $\eta$ opacity due to electrons associated with baryons
increases and eventually dominates over pair opacity. Straightforward
calculations with the laboratory density profile (\ref{nxi}) gives
\begin{equation}
\tau=\left\{
\begin{array}
[c]{cc}%
\dfrac{1}{6}\tau_0\left(\dfrac{R_0}{R}\right)^3, & R_0\ll R \ll\eta R_0,\\
& \\
\dfrac{1}{2\eta^2}\tau_0\left(\dfrac{R_0}{R}\right), & \eta R_0\ll R\ll\eta^2 l,\\
& \\
\tau_0\dfrac{R_0 l}{R^2}, & R\gg\eta^2 l,
\end{array}
\right.
\label{tau2}%
\end{equation}
where%
\begin{equation}
\tau_{0}=\sigma n_{0}R_{0}=\frac{\sigma E_0}{4\pi m_p c^2 R_0 l \eta}
=\frac{\sigma L}{4\pi m_p c^3 R_0 \eta}.
\label{tau0}%
\end{equation}

Let us interpret the formula (\ref{tau2}). On the one hand, first two lines in
(\ref{tau2}) correspond to the case when the photon propagates inside the
outflow for a significant time so that the number density on its path
substantially decreases before it leaves. In this respect the outflow is ``a
long wind'', even if the laboratory thickness of the outflow may be small, $l\ll
R$. We refer to this case as a \emph{photon thick outflow}. On the other hand,
the third line in (\ref{tau2}) corresponds to the case when the number density
of the outflow does not change substantially on the photon world line before it
escapes. In this respect the outflow is ``a thin shell'' even if the duration
of explosion could be long and a wind was launched. We refer to this latter
case as a photon thin outflow. For instance, a geometrically thin
ultrarelativistically expanding shell may be both thin or thick with respect to
the photon propagating inside it.

Similar consideration may be applied to a photon emitted at any distance
$\xi$ from the outer boundary of the outflow, see Section \ref{geomdyn}. It is
clear then, that even in a photon thick outflow there is always a photon thin
layer located near the outer boundary. During acceleration phase such a photon
thin part accounts for a fraction not larger than $\Gamma^{-1}$ of the entire
width of the outflow\footnote{Formally photon thin accelerating solution
exists, and it is given by the last line in (\ref{tau2}). However, its validity
condition is $l\ll R_0^2/R=R_0/\Gamma$.}.

In the derivation of (\ref{tau2}) we considered a portion of a relativistic
wind, but these results are generic and apply to any density profile of the outflow.

\begin{table*}[t]
\caption{Dependencies of transparency parameters on initial
conditions: energy $E_{0}$, entropy $\eta$, and radius $R_{0}$ for the shell model.}%
\label{shelldependence}%
\center \begin{tabular}
[c]{l|lll|lll|lll|lll}%
Regime of transparency & \multicolumn{3}{c|}{$R_{ph}$} &
\multicolumn{3}{c|}{$\Gamma_{ph}$} & \multicolumn{3}{c|}{$kT_{ph}$} &
\multicolumn{3}{c}{$kT_{obs}$}\\\hline Pair & $E_{0}^{1/4}$ &  & $R_{0}^{1/4}$
& $E_{0}^{1/4}$ &  & $R_{0}^{-3/4}$ &
\multicolumn{3}{c|}{$0.040m_{e} c^{2}$} & $E_{0}^{1/4}$ &  & $R_{0}^{-3/4}$\\
Acceleration & $E_{0}^{1/3}$ & $\eta^{-1/3}$ & $R_{0}^{1/3}$ & $E_{0}^{1/3}$ &
$\eta^{-1/3}$ & $R_{0}^{-2/3}$ & $E_{0}^{-1/12}$ & $\eta^{1/3}$ &
$R_{0}^{-1/12}$ &
$E_{0}^{1/4}$ &  & $R_{0}^{-3/4}$\\
Coasting photon thick & $E_{0}$ & $\eta^{-3}$ & $R_{0}^{-1}$ &  & $\eta$ &  &
$E_{0}^{-5/12}$ & $\eta^{5/3}$ & $R_{0}^{7/12}$ & $E_{0}^{-5/12}$ &
$\eta^{8/3}$ &
$R_{0}^{7/12}$\\
Coasting photon thin & $E_{0}^{1/2}$ & $\eta^{-1/2}$ &  &  & $\eta$ &  &
$E_{0}^{-1/12}$ &  & $R_{0}^{-1/12}$ & $E_{0}^{-1/12}$ & $\eta$ &
$R_{0}^{-1/12} $%
\end{tabular}
\end{table*}

The \emph{photospheric radius} $R_{ph}$ is defined by equating (\ref{tau2}) to
unity. It is worth noting that it is the lower limit in the integral
(\ref{tau}) that is associated with the photospheric radius, but not the upper
one. The upper limit in (\ref{tau}) is the radius at which the photon leaves
the outflow, even if it may decouple from the outflow at much smaller radius,
as in the photon thick case. The characteristic parameter which determines the
type of the outflow at the photospheric radius is the ratio
$\tau_0/(4\Gamma^4)$. When it is much smaller than one the outflow is photon
thick, while in the opposite case the outflow is photon thin.

Consider now the shell model of GRBs with typical parameters expressing their total
energy as $E_{0}=10^{54}E_{54}$ erg, initial size as $R_{0}=l=10^{8}R_{8}$ cm
and entropy parameter as $\eta =10^{2}\eta_{2}$. We find the following asymptotic
solutions for the photospheric radius together with domains of their applicability
%\begin{onecolumn}%
\begin{equation}
R_{ph}=\left\{
\begin{array}
[c]{lc}%
4.4\times10^{10}\left(  E_{54}R_{8}\right)  ^{1/4}\text{ cm}, & \\
& \\
\hfill E_{54}\ll4.8\times10^{-20}\eta_{2}^{4}R_{8}^{-1}, & \\
& \\\hline
& \\
1.8\times10^{12}\left(  E_{54}\eta_{2}^{-1}R_{8}\right)  ^{1/3}\text{ cm}, &
\\
& \\
\hfill4.8\times10^{-20}\eta_{2}^{4}R_{8}^{-1}\ll E_{54}\ll3.2\times10^{-8}%
\eta_{2}^{4}R_{8}^{2}, & \\
& \\\hline
& \\
1.8\times10^{17}E_{54}%
\eta_{2}^{-3}R_{8}^{-1}\text{ cm}, & \\
& \\
\hfill3.2\times10^{-8}\eta_{2}^{4}R_{8}^{2}\ll E_{54}\ll1.1\times10^{-5}%
\eta_{2}^{5}R_{8}^{2}, & \\
& \\\hline
& \\
5.9\times10^{14}\left(  E_{54}\eta_{2}^{-1}\right)  ^{1/2}\text{ cm}, & \\
& \\
\hfill E_{54}\gg1.1\times10^{-5}\eta_{2}^{5}R_{8}^{2}. &
\end{array}
\right. \label{shell}%
\end{equation}
%\end{onecolumn}

\begin{figure}[ptb]%
\centering
\includegraphics[width=\columnwidth]{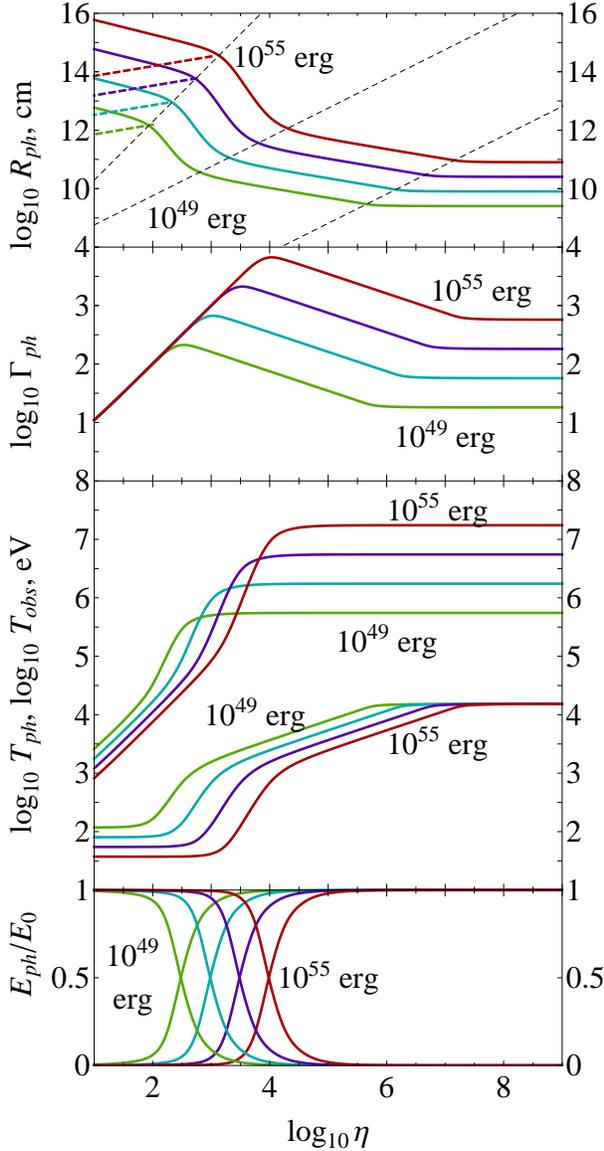}
\caption{From top to bottom: photospheric radius $R_{ph}$, Lorentz factor $\Gamma_{ph}$,
observed $T_{obs}$ and comoving $T_{ph}$ temperatures,
and ratio of the energy emitted to the total energy $E_{Ph}/E_0$
at photospheric radius as functions of entropy $\eta$ for shells with different
total energy $E_0$ but the same width $l=R_0=10^8$ cm.
All four regimes with different asymptotics are clearly visible
and dashed black lines corresponding to their domain of validity from
Eq. (\ref{shell}) are shown. Curves are drawn for $E_0$ equal to:
$10^{49}$~erg (green), $10^{51}$~erg (blue), $10^{53}$~erg (violet),
and $10^{55}$~erg (red). Dashed thick lines denote the diffusion radius for each energy.
\label{full}}
\end{figure}

For very small baryonic loading, or in other words for a pure electron-positron
plasma, the photospheric radius does not depend on $\eta$ parameter. For
increasing baryonic loading it increases as $\eta^{-1/3}$ (accelerating photon
thick solution). In both these cases the Lorentz factor at the photosphere is
not equal to $\eta$, but it is much smaller. For larger baryonic loading the
photospheric radius steeply increases as $\eta^{-3}$ (coasting photon thick
solution), and finally it increases as $\eta^{-1/2}$ (coasting photon thin
solution), see Fig. \ref{full}. There we also show as function of the entropy
parameter the following quantities computed at the photospheric radius: the
Lorentz factor, the observed and comoving temperatures, and fraction of energy
emitted from the photosphere to the total energy, for different values of the
total energy $E_{0}$. Dependence of parameters of transparency on initial
conditions is also illustrated in Tab. \ref{shelldependence}. It is clear that
the highest Lorentz factors at photospheric radius are attained in photon thick
asymptotics. The largest transparency radii are reached instead in photon thin
asymptotics.

%\clearpage

\begin{figure}[ptb]%
\centering
\includegraphics[width=\columnwidth]{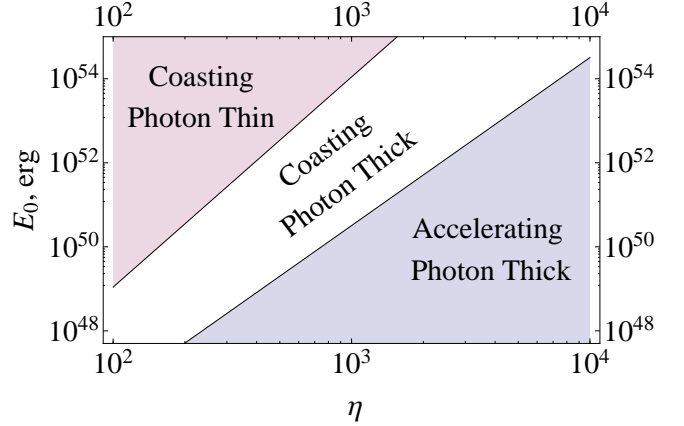}
\caption{The energy-baryonic loading diagram showing the validity of
the various asymptotic solutions for the photospheric radius for typical
parameters of GRBs with $l=R_0=10^8$ cm.
\label{EB}}
\end{figure}

Fig. \ref{EB} shows the energy-baryonic loading diagram, where the regions of
validity of the asymptotics discussed above are indicated explicitly for
typical parameters of GRBs. For all the relevant range of GRBs parameters
$10^{48}$ erg${}<E_{0}<10^{55}$ erg and $10^{6}$ cm${}<R_{0}<10^{12}$ cm all
four asymptotics are present in the interval $10<\eta<10^{10}$.

In the case of gradual energy release resulting in relativistic wind an
additional parameter is present, that is the duration of energy release, which
we parameterize as $\Delta t=1 \Delta t_{1}$~s. Instead of the total energy
$E_{0}$ the luminosity $L=10^{50} L_{50}$ erg/s is used. The corresponding
photospheric radius is
\begin{equation}
R_{ph}=\left\{
\begin{array}
[c]{lc}%
8.1\times10^{8}L_{50}^{1/4}R_{8}^{1/2}\text{ cm}, & \\
& \\
\hfill L_{50}\ll5.3\times10^{-15}\eta_{2}^{4}R_{8}^{-2}, & \\
& \\\hline
& \\
1.3\times10^{10}\left(  L_{50}\eta_{2}^{-1}R_{8}^{2}\right)  ^{1/3}\text{
cm}, & \\
& \\
\hfill5.3\times10^{-15}\eta_{2}^{4}R_{8}^{-2}\ll L_{50}\ll9.8\times10^{-2}%
\eta_{2}^{4}R_{8}, & \\
& \\\hline
& \\
5.9\times10^{10}L_{50}\eta_{2}%
^{-3}\text{ cm}, & \\
& \\
\hfill9.8\times10^{-2}\eta_{2}^{4}R_{8}\ll L_{50}\ll10^{5}\eta_{2}^{5}\Delta
t_{1}, & \\
& \\\hline
& \\
5.9\times10^{10}\left(  L_{50}\Delta t_{1}\eta_{2}^{-1}\right)  ^{1/2}\text{
cm}, & \\
& \\
\hfill L_{50}\gg10^{5}\eta_{2}^{5}\Delta t_{1}. &
\end{array}
\right. \label{wind}%
\end{equation}

\begin{table*}[ptb]
\caption{Dependencies of transparency parameters on initial
conditions: energy release rate $L$, entropy $\eta$, duration $\Delta t$, and
radius $R_{0}$ for the wind model.}%
\label{winddependence}%
\center \begin{tabular}
[c]{l|lll|lll|lll|llll}%
Regime of transparency & \multicolumn{3}{c|}{$R_{ph}$} &
\multicolumn{3}{c|}{$\Gamma_{ph}$} & \multicolumn{3}{c|}{$kT_{ph}$} &
\multicolumn{4}{c}{$kT_{obs}$}\\\hline Pair & $L^{1/4}$ & & $R_{0}^{1/2}$ &
$L^{1/4}$ &  & $R_{0}^{-1/2}$ &
\multicolumn{3}{c|}{$0.040m_{e} c^{2}$} & $L^{1/4}$ & & & $R_{0}^{-1/2}$\\
Acceleration & $L^{1/3}$ & $\eta^{-1/3}$ & $R_{0}^{2/3}$ & $L^{1/3}$ &
$\eta^{-1/3}$ & $R_{0}^{-1/3}$ & $L^{-1/12}$ & $\eta^{1/3}$ & $R_{0}^{-1/6}$ &
$L^{1/4}$ & & & $R_{0}^{-1/2}$\\
Coasting photon thick & $L$ & $\eta^{-3}$ &  &
  & $\eta$ &  &
$L^{-5/12}$ & $\eta^{5/3}$ & $R_{0}^{1/6}$ &
$L^{-5/12}$ & $\eta^{8/3}$ & & $R_{0}^{1/6}$\\
Coasting photon thin & $L^{1/2}$ & $\eta^{-1/2}$ & $\Delta t^{1/2}$ &
 & $\eta$ & &
$L^{-1/12}$ & $\Delta t^{-1/3}$ & $R_{0}^{1/6}$ & $L^{-1/12}$ & $\eta$ &
$\Delta t^{-1/3}$ &
$R_{0}^{1/6}$%
\end{tabular}
\end{table*}

\begin{figure}[ptb]%
\centering
\includegraphics[width=\columnwidth]{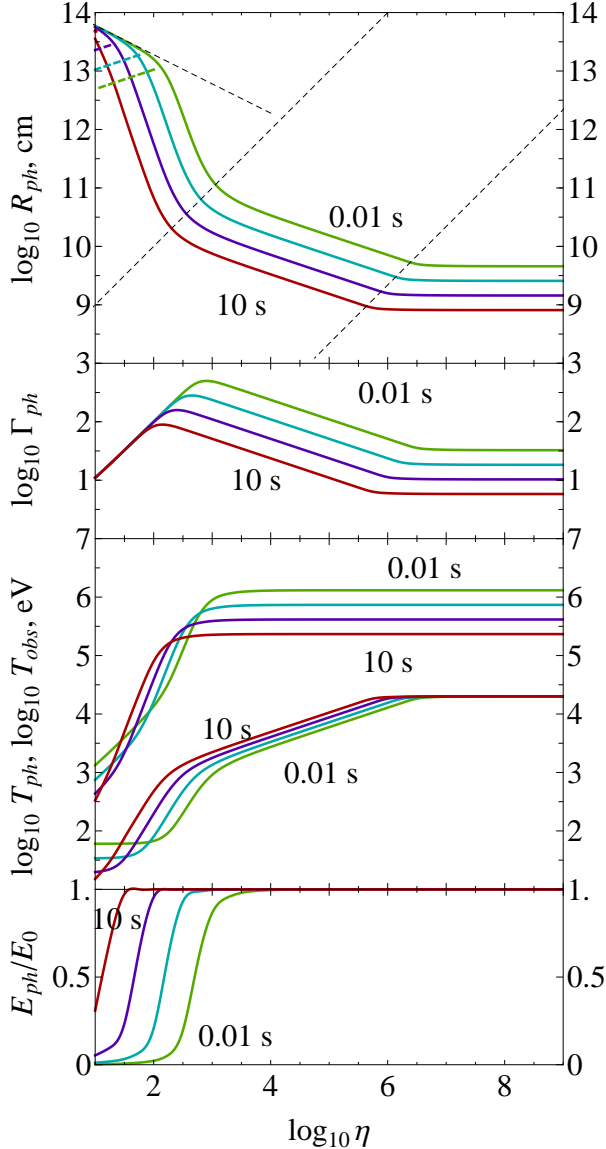}
\caption{The same as in Fig. \ref{full} for winds with different
duration, but the same total energy $E_0=10^{51}$ erg and radius of origin $R_0=10^8$ cm.
All four regimes with different asymptotics are clearly visible.
Curves are drawn for $\Delta t$ ranging from $10^{-2}$~s (green)
to $10$~s (red) in steps of one order of magnitude.
\label{fullw}}
\end{figure}

\begin{figure}[ptb]%
\centering
\includegraphics[width=\columnwidth]{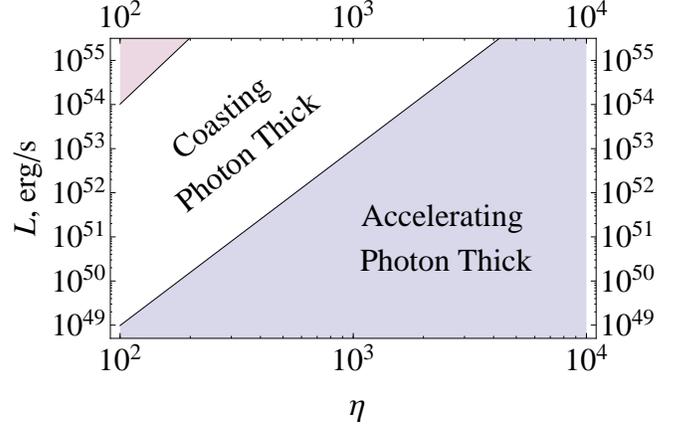}
\caption{The luminosity-baryonic loading diagram showing the validity of
the various asymptotic solutions for photospheric radius of wind
with duration $\Delta t=0.1$~s.
Notation is the same as on Fig. \ref{EB}.
\label{EBw}}
\end{figure}

In Fig. \ref{fullw} we show as function of the entropy parameter the
following quantities computed at the photospheric radius: the Lorentz factor,
the observed and comoving temperatures, and fraction of energy emitted from
the photosphere to the total energy, for different duration of the wind with
the total energy $E_{0}=10^{51}$~erg, and inner boundary radius $R_{0}=10^{8}%
$~cm. Wind duration ranges from $10$ ms to $10$ s. The corresponding wind
luminosity varies from $10^{53}$ erg/s to $10^{50}$ erg/s. Fig. \ref{EBw}
shows the luminosity-baryonic loading diagram where the regions of validity of
the asymptotics discussed above are indicated. Dependence of
transparency parameters on the initial conditions of the wind is presented in
Tab. \ref{winddependence}.

\subsection{Comparison with previous works}

The expressions for the optical depth of a relativistic wind were obtained e.g.
by \citet{2000ApJ...530..292M}. Their formulas coincide with our (\ref{tau2})
up to a numerical factor which comes from the integration over the radial
coordinate. It should be noted, however, that only the photon thick asymptotic
limit is discussed in \citet{2000ApJ...530..292M}. The photon thin asymptotic
limit may also be valid for relativistic winds in the coasting phase, provided
that $l\ll R/(2\eta^{2})$. This is an independent condition from $\Delta t\gg
R_{0}/c$ and it is therefore possible to give the following constraints for
$\Delta t$ under which the outflow takes the form of a wind, but it is photon
thin at the photospheric radius:
\begin{equation}\label{deltaTphotonthin}
    \frac{R_{0}}{c}\ll\Delta t\ll\dfrac{\tau_0}{4\eta^4}\frac{R_{0}}{c}.
\end{equation}
photospheric radius for photon thick and photon thin asymptotics for a portion
of the coasting relativistic wind was obtained by \citet{2002MNRAS.336.1271D}.

Similar considerations apply to an ultrarelativistic shell which is considered
e.g. in \citet{1990ApJ...365L..55S} and \citet{1993ApJ...415..181M} and by
\citet{2000A&A...359..855R} in the photon thin approximation. The corresponding
condition that the shell at the photospheric radius appears to be photon thin
is $\tau_0\gg 4\eta^4$. It is possible, however, that initial conditions
satisfy the opposite constraint which results in a photon thick shell, see
\citet{2005ApJ...635..516N}.

Numerical hydrodynamic simulations produce complex density, temperature and
Lorentz factor profiles of the outflow. In particular,
\citet{1993MNRAS.263..861P} and \citet{1993ApJ...415..181M} considered an
explosion in a compact region with radius $R_{0}$\ and studied numerically the
hydrodynamic evolution of an optically thick plasma with various initial
conditions. They have shown that the plasma forms a relativistically expanding
shell with some density and velocity profiles. The characteristic width of the
density profile appears to be constant up to large radii, but later it
increases linearly with radius due to the fact that the Lorentz factor appears
to be monotonically increasing within the main part of the profile
\begin{align}
    l & \simeq R_{0}, & R & <\eta^{2}R_{0},\label{constthick}\\
    l & \propto R,    & R & >\eta^{2}R_{0}.\label{nonconstthick}%
\end{align}
Such a spreading in density profile may result in a substantial increase in the
width $l\gg R_{0}$ of the shell when it becomes transparent to radiation. Note
that condition (\ref{nonconstthick}) coincides with the condition in the third
line of (\ref{tau2}), which corresponds to the case of a photon thin outflow.
This coincidence may give an impression that photon thin asymptotics does not
exist for hydrodynamically spreading outflows. Nevertheless direct integration
of (\ref{tau}) in numerical simulations shows that the photon thin asymptotics
is valid even for the shell undergoing such spreading.

All asymptotic solutions for the optical depth have been considered by
\citet{2002ApJ...578..812M}, except the case of pure electron-positron outflow.
They derived the photospheric radius considering the expansion in comoving
reference frame. Notice, that the photon thin asymptotics is obtained in
\citet{2002ApJ...578..812M} by assuming hydrodynamic spreading of the outflow
found by the same authors in \citep{1993ApJ...415..181M}. In absence of such
spreading (e.g. portion of wind considered above) such asymptotics cannot be
obtained this way. Finally, \cite{2011MNRAS.415.1663T} discuss all asymptotic
solutions, applying them to a relativistic wind.

We conclude that interpretation of the formula (\ref{tau2}) it terms of photon
thick and photon thin conditions given above provides additional physical
insight to the consideration of optical depth in shell and wind models.

\section{Photon diffusion}
\label{diffusionSec}

Up to now we assumed that photons escape the expanding outflow when its
optical depth decreases to unity. We have distinguished two possibilities:

\begin{itemize}
\item in the photon thick case electron number density decreases along the
photon path so rapidly that the medium becomes too rarified to sustain
collisions. Most photons however still remain inside the outflow after
decoupling;

\item in the photon thin case the variation of the electron number density
along the photon path can be neglected, but the mean free path of photons
increases with expansion and eventually exceeds the radial thickness of the
outflow.
\end{itemize}

In all the previous discussion we explicitly neglected the effect of radiative
diffusion out of the outflow by random walks of photons. Diffusion time is
given by $t_{D,c}=l_{c}^{2}/D_c$, where $l_{c}=\Gamma l$ is the comoving radial
thickness of the outflow, and diffusion coefficient is $D_{c}=(c\lambda
_{c})/3=c/(3\sigma n_{c})$, where $\lambda _{c}$ and $n_{c}$\ are comoving mean
free path of photons and comoving electron number density, respectively.

In order to determine at which radii diffusion becomes important one has to
compare this diffusion time with comoving expansion time of the outflow
$t_{c}=R/(c\Gamma)$. Taking into account (\ref{nxi}) and (\ref{tau0}) we obtain
that it happens when the outflow reaches the radius
\begin{equation}\label{diffradius}
    R_D=\left(\tau_{0}\eta^{2}R_0 l^2\right)^{1/3}\simeq
    \begin{cases}
    7.2\times 10^{13} (E_{54}l_8\eta_2)^{1/3} \text{ cm}, \\
    5.0\times 10^{11} (L_{50}\Delta t_1^2\eta_2)^{1/3} \text{ cm}.
    \end{cases}
\end{equation}
This diffusion radius turns out to be always larger than the photospheric
radius of photon thick outflows, $R_D\gg R_{ph}$, so that diffusion is
irrelevant for their description. In the opposite case of photon thin outflows
instead the diffusion radius is always smaller than the photospheric radius
$R_D\ll R_{ph}$. In this case most radiation leaves the photon thin outflow not
at its photospheric radius, but before it reaches the diffusion radius, when
the outflow is still opaque. In other words, the decoupling of photons from the
outflow occurs not locally, as in the photon thick case, but near its
boundaries where photons are transferred to by diffusion. In this sense the
characteristic radius of the photospheric emission is not the photospheric
radius found from (\ref{tau2}), but the radius of diffusion (\ref{diffradius}).
Besides, the comoving temperature of escaping radiation is different from that
discussed in Section \ref{optdepth}.

In what follows we consider decoupling of photons from photon thick and photon
thin outflows separately.

\section{Photospheric emission from photon thick outflows}

\subsection{Geometry and dynamics of the photosphere}
\label{geomdyn}

Unlike traditional static sources usually dealt with in astrophysics,
relativistic outflows may have strongly time-varying photospheres.
For the portion of wind the optical depth can be calculated
analytically both at acceleration and coasting phases for photon thin and
photon thick outflows. The result is
\begin{multline}\label{tauangle}
    \tau(r,\theta,t)=\tau_{0}R_{0}\Biggl\{\frac{1}{r\sin\theta}\left[  \theta
        -\tan^{-1}\left(  \dfrac{r\sin\theta}{cT+r\cos\theta}\right)  \right]
\\
        -\beta_m  \left(  \dfrac{1}{r}-\dfrac
        {1}{\sqrt{(cT+r\cos\theta)^{2}+(r\sin\theta)^{2}}}\right)
\\
        +\frac{R_{0}^{2}}{6}\left(  \frac{1}{r^{3}}-\frac{1}{\left[  (cT+r\cos
        \theta)^{2}+(r\sin\theta)^{2}\right]  ^{3/2}}\right)  \Biggr\},
\end{multline}
where $T$ is the time interval during which photon remains inside the outflow,
determined by the equations of motion of the photon and of the outflow, and
$\beta_m=1-1/(2\eta^{2})$. For a given laboratory time $t$ the photosphere
geometry $r=r(\theta)$ is obtained by equating (\ref{tauangle}) to unity. Then
formula (\ref{tauangle}) gives complete information on the dynamics and
geometry of the photosphere of portion of ultrarelativistic wind. In order to
understand this dynamics it is instructive to consider its limiting cases.

\begin{figure}[ptb]%
\centering
\includegraphics[
width=\columnwidth
]%
{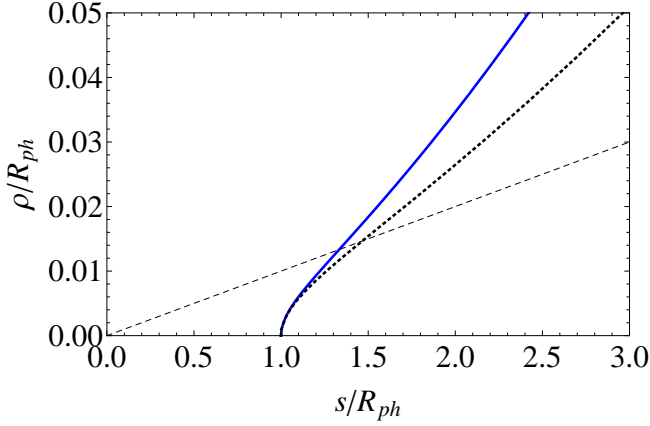}%
\caption{The shape of photospheres of infinitely long coasting
(blue solid curve) and accelerating (black dotted curve) winds
for $\Gamma_{ph}=100$. Dashed line shows the relativistic beaming angle.
\label{WindShell}}
\end{figure}

Firstly, the photosphere of the coasting infinitely long relativistic wind with
$\Gamma={}$const analyzed by \citet{1991ApJ...369..175A} may be recovered from
(\ref{tauangle}) with $T\rightarrow\infty$. In that case the last term in
(\ref{tauangle}) can be neglected and we have (see e.g. \citet{2008ApJ...682..463P})
\begin{equation}\label{PhotosphereInfWindCoasting}
    \frac{r}{R_{0}}=\tau_{0}\Biggl(\frac{\theta}{\sin\theta}-\beta_m\Biggr),
\end{equation}
which is a static surface having concave shape, see Fig. \ref{WindShell}.

Secondly, the photosphere of the accelerating infinite wind may be obtained
from (\ref{tauangle}) for $T\rightarrow\infty$ and $\eta\rightarrow\infty$. It
results in a qubic equation describing a static surface with curvature larger
than that of the coasting wind, see Fig. \ref{WindShell}. In both cases these
photospheres appear for a distant observer as static spots with radius
\begin{equation}
\rho=\pi\tau_{0}R_{0}\label{WindSpot}%
\end{equation}
and brightness decreasing from the center to the edge.

Now consider dynamic properties of the photosphere of photon thick outflow
described by (\ref{tauangle}) as seen by a distant observer. The arrival time
of radiation is defined as $t_{a}=t-r\cos\theta/c$. The equitemporal surface
(EQTS) of the photospheric emission represents a part of the photosphere
visible at a given instant of arrival time $t_{a}$, see e.g.
\citet{2001ApJ...555L.113R}. We will refer to that surface as
\emph{Photospheric EQTS} (PhE).

\begin{figure}[ptb]
\centering
\includegraphics[
width=\columnwidth
]{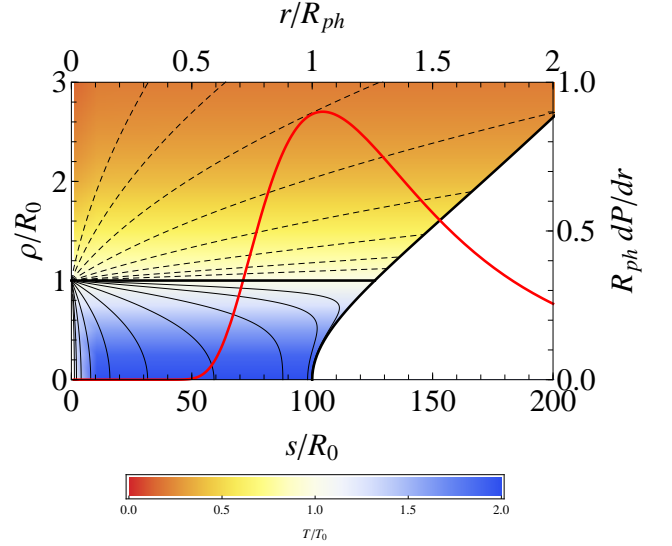}
%\\\centering
%\includegraphics[
%height=2.6091in,
%width=2.7985in
%]%
%{talumwind}
\caption{PhE of the photon thick accelerating outflow at different
arrival times, and the probability density function of photon last
scattering integrated over angles $P(r)$ (red thick curve).
Thick black curve represents the photosphere of infinitely long
accelerating wind. PhEs are illustrated for several arrival times
with logarithmic spacing by thin black curves. The surface $\cos\theta=\beta$
is given by $\rho=R_{0}$ and it is shown by thick black line.
Dashed curves illustrate the maximal visible size $\rho$ for several arrival
times with logarithmic spacing. The PhE at that arrival times is a
part of the wind photosphere limited by the corresponding curves.
Range of observed temperature of emission
under the asymptotic photosphere is illustrated by color, see legend. Here
$R_{ph}=100R_{0}$.}%
\label{AccelWindAppPhotosphere}%
\end{figure}

\begin{figure}[ptb]%
\centering
\includegraphics[
width=\columnwidth
]%
{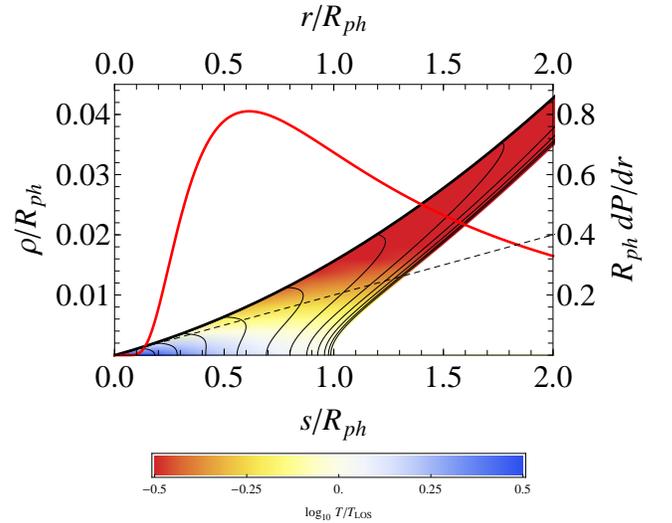}%
\caption{The same as in Fig. \ref{AccelWindAppPhotosphere}
for a photon thick coasting outflow.
PhEs are illustrated by thin curves for several arrival
times with linear spacing. Thick black curve bounding PhEs
correspond to the position of maximal visible angles at given $t_a$.
Notice that these angles exceed the relativistic beaming angle,
shown by dashed black line. Right edge of the colored area is
the photosphere of infinitely long coasting wind
(\ref{PhotosphereInfWindCoasting}). Here $\Gamma=100$.}
\label{WindAppPhotosphere}
%\label{blongwind}}
\end{figure}

PhE of the photon thick outflow has concave shape, see Fig.
\ref{AccelWindAppPhotosphere} and Fig. \ref{WindAppPhotosphere} for
accelerating and coasting cases, respectively. This concave PhE in both cases
approaches the photosphere of infinitely long wind. In the coasting case the
approach to that surface is only asymptotic, while in the accelerating case the
photosphere actually reaches it at finite arrival time. The external boundary
of the PhE for a given $t_{a}$ shown in Fig. \ref{WindAppPhotosphere} is
defined by the condition that the optical depth for photons emitted from the
outermost layer of the outflow equals unity. Notice that this boundary is wider
than the relativistic beaming surface (these are tube and cone for accelerating
and coasting outflows, respectively). As soon as the innermost part of the
outflow reaches the photospheric radius, i.e. observer sees the switching off
of the wind, the inner boundary of the PhE expands with $t_a$. The surface of
these boundaries is given by (\ref{PhotosphereInfWindCoasting}) in the case of
coasting photon thick outflow.

\subsection{Observed flux and spectrum}
\label{obsflux}

The basis of spectrum and flux calculation is radiative transfer equation for
specific intensity $I_\nu$ along the ray (see e.g. \cite{1979rpa..book.....R},
p.~11)
\begin{equation}\label{radtransf}
    \frac{dI_\nu}{ds}=j_\nu-\kappa_\nu I_\nu,
\end{equation}
where $j_\nu$ is monochromatic emission coefficient, $\kappa_\nu$ is absorption
coefficient and $s$ is distance, measured along the ray.

Spectral intensity of radiation at infinity on a ray coming to observer at some
arrival time $t_a$ is given by formal solution of this equation
\begin{multline}\label{sourcefunc}
    I_\nu(\nu,\rho,t_a)=\int \mathcal{I}_\nu(\nu,r,\theta,t)
        \frac{d}{ds}\left\{\exp[-\tau(\nu,r,\theta,t)]\right\}\, ds\\
    =\int \mathcal{I}_\nu(\nu,r,\theta,t)
        \exp[-\tau(\nu,r,\theta,t)]\, d\tau,
\end{multline}
where $\mathcal{I}_\nu(r,\theta,t)$ is the source function, equal to the ratio
of emission and absorption coefficients $\mathcal{I}_\nu=j_\nu/\kappa_\nu$,
optical depth $\tau$ is an integral of $\kappa_\nu$ from the point on the ray
under consideration to infinity $\tau=\int\kappa_\nu ds$, and variables
$(r,\theta,t)$ are connected by $t_a=t-r\cos\theta$ and $r\sin\theta=\rho$, see
Fig. \ref{xi}. We use Thomson scattering cross section in comoving frame
$\kappa_c=const$.

Total observed flux is an integral over all rays
\begin{equation}\label{Fnu}
    F_\nu(\nu,t_a)=2\pi\Delta\Omega \int \rho\, d\rho\, I_\nu(\nu,\rho,t_a)\,
\end{equation}
where $\Delta\Omega$ is the solid angle of the observer's detector as seen from
the outflow in the laboratory frame and $2\pi \rho d \rho$ is an element of area in the
plane of the sky.

In what follows we assume that emissivity $j_\nu$ is thermal and isotropic in
comoving frame of the outflow. The laboratory source function is then
\begin{equation}%\label{}
    \mathcal{I}_\nu(\nu,r,\theta,t)=\frac{2h}{c^2}\frac{\nu^3}{
        \exp\left(\frac{h\nu \Gamma(1-\beta\cos\theta)}{kT_c(r,t)}\right)-1}.
\end{equation}
This approximation is justified when the radiation field is tightly coupled to
the matter. The photospheric emission comes from entire volume of the outflow, and the
computational method sketched above is closely related to that used in
\citep{2011ApJ...737...68B} where the concept of ``fuzzy photosphere'' was
introduced. This method will be referred to as \emph{fuzzy photosphere}
approximation.

Most of energy reaching observer is emitted from the region near the PhE,
where the probability density function along the ray
\begin{equation}\label{PDF}
    P(r,\theta,t)=P_0 \frac{d}{ds}\exp[-\tau(r,\theta,t)]
\end{equation}
with $P_0$ being normalization, reaches the maximum. For this reason the
dynamics of PhE studied in the previous section determines both light curves
and spectra of observed photospheric emission. When the time dependence in this
equation is discarded this $P(r,\theta)$ coincides with the probability density
function of the last scattering defined in \citep{2008ApJ...682..463P}.
Assuming that all the energy comes from the PhE only, i.e. a surface instead of
the volume discussed above, the computation may be reduced to one dimensional
integration by substitution of the function $P$ with a Dirac delta. Such more
crude approximation, in contrast to the fuzzy photosphere one, will be referred
to as \emph{sharp photosphere} approximation.

For photon thick outflow the optical depth becomes function of $r$ and $\theta$
only and the comoving temperature also depends only on radius. In this respect
the photon thick case is similar to the infinite wind. Then the integrand in
(\ref{Fnu}) does not depend on time and only limits of integration provide time
dependence due to motion of the outflow boundaries. The probability density
function (\ref{PDF}) integrated over angles is shown in Figs.
\ref{AccelWindAppPhotosphere} and \ref{WindAppPhotosphere} for accelerating and
coasting photon thick outflows.

\begin{figure}[ptb]
\centering
\includegraphics[width=\columnwidth]{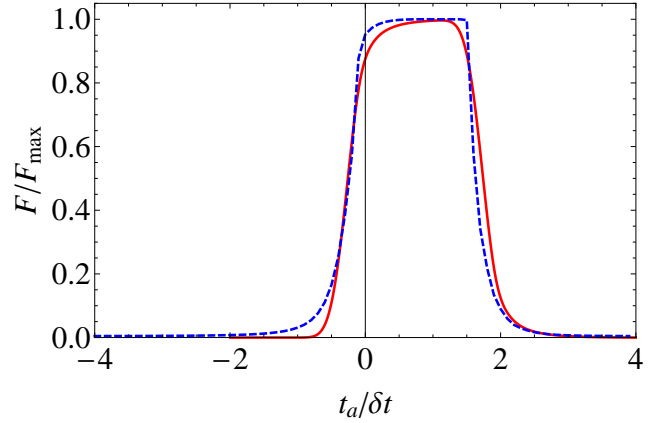}
\caption{The light curve of photospheric emission
from the photon thick accelerating outflow
in fuzzy (red curve) and sharp photosphere approximations (dashed blue curve).
Here $R_{ph}=100R_{0}$, and $\Delta t=2\delta t$, see (\ref{deltat}) below.}
\label{AccelWindLC}
\end{figure}

The observed flux of photospheric emission from accelerating outflow is
illustrated in Fig. \ref{AccelWindLC} by thick red curve (fuzzy photosphere)
and by dotted blue curve (sharp photosphere). The characteristic raising and
decaying time is in both cases
\begin{equation}\label{deltat}
    \delta t=\frac{R_{0}^{2}}{(R_{ph}c)}=\frac{R_0}{\Gamma_{ph}c}.%
\end{equation}
There is no simple analytic expression describing full light curve, however its
decreasing part is close to a power law with index $-4.7$ and $-6.5$ within
fuzzy and sharp photosphere approximations, respectively. As minimal
duration of the photon thick outflow $\Delta t$ is of order $R_0/c$, then
$\Delta t\gg\delta t$ and the light curve has almost rectangular shape.

Such accelerating outflow appears to a distant observer as a spot with size
$\rho=(R_{0}^{2}-(t_{a}/c)^{2})^{1/2}$, for $-R_{0}/c\leq t_{a}\leq0$. As soon
as the PhE reaches the corresponding accelerating infinitely long wind
photosphere at $t_{a}=0$ the spot size starts to increase almost linearly with
time $\rho\simeq R_{0}+ct_{a}$. Finally, as the innermost part of the outflow
reaches the photospheric radius the spot transforms to a ring with rapidly
decreasing width and brightness.

\begin{figure}[ptb]
\centering
\includegraphics[width=\columnwidth]{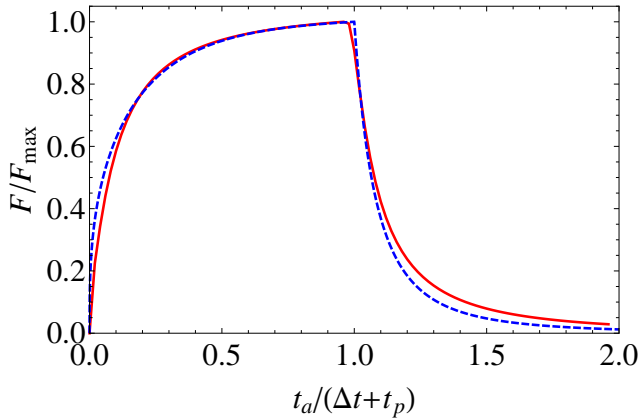}
\caption{The same as in Fig. \ref{AccelWindLC}
for a photon thick coasting outflow.
Here $\Gamma=100$ and $\Delta t=5t_p$, see (\ref{tpcoastthin}) below.}
\label{WindLC}
\end{figure}

The observed photospheric emission of the coasting photon thick outflow
results
in the flux changing as
\begin{equation}
    F=F_{max}\left[1-(t_{p}/t_{a})^{2}\right] ,
\end{equation}
with
\begin{equation}\label{tpcoastthin}
    t_p=\frac{R_{ph}}{2\eta^2 c},
\end{equation}
i.e. increase up to the saturation value $F_{max}\propto L$, see the raising
part of the light curve in Fig. \ref{WindLC}, both in sharp and fuzzy photosphere
approximations. Radius of the visible spot then reaches its maximal
size (\ref{WindSpot}). As arrival time exceeds $t_{p}+\Delta t$ the innermost
part of the outflow approaches the wind photosphere
(\ref{PhotosphereInfWindCoasting}) along the line of sight and the spot
transforms to a ring, the flux decreases rapidly in both approximations
\begin{equation}\label{swithchoff}
    F\propto t_{p}^{2}\left[ \frac1{(t_{a}-\Delta
        t)^{2}}-\frac1{t_{a}^{2}}\right].
\end{equation}
For $t_{a}\gg\Delta t$ it behaves as $F\propto t_{a}^{-3}$, see the
decreasing part of the light curve in Fig. \ref{WindLC}. Similarly
to the accelerating photon thick outflow the light curve for $\Delta t\gg
t_{p}$ has almost rectangular shape due to the fact that its increase and decay
times are much shorter than $\Delta t$.


\begin{figure}[ptb]%
\centering
\includegraphics[
width=\columnwidth
]%
{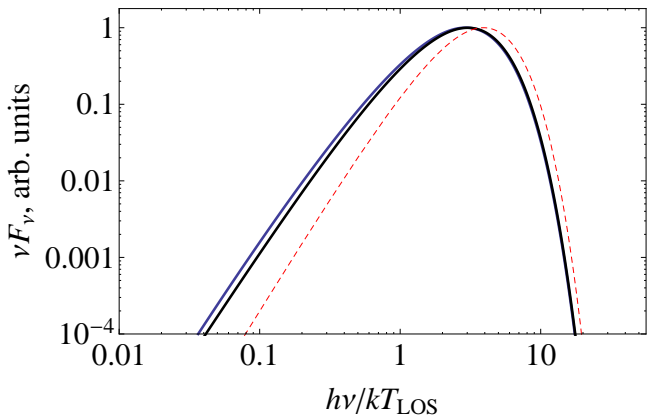}%
\caption{Instantaneous spectrum of photospheric emission of
accelerating photon thick outflow in continuous (blue thick curve)
and sharp photosphere approximation (black thick curve).
Dashed red curve represents the thermal spectrum with the temperature at the
line of sight $T_{LOS}$. Lorentz factor at photospheric radius is $\Gamma_{ph}=100$.
\label{AccelWindSpectrum}}
\end{figure}

Accelerating photon thick outflows exhibit photospheric spectra close to
thermal ones, see Fig. \ref{AccelWindSpectrum}. In ultrarelativistic case
spectra computed using both sharp and fuzzy photosphere approximations are
very close to each other. Both have small deviations from thermal spectrum in
the low energy part with the corresponding Band low energy indices
$\alpha=0.82$ and $\alpha=0.71$, respectively.

\begin{figure}[ptb]%
\centering
\includegraphics[
width=\columnwidth
]%
{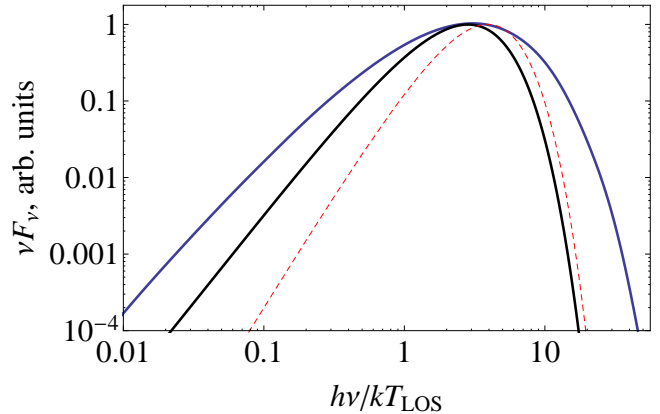}%
\caption{The same as in Fig. \ref{AccelWindLC}
for a photon thick coasting outflow with Lorentz factor $\Gamma=100$.
\label{CoastWindSpectrum}}
\end{figure}

In contrast, the spectrum of photospheric emission of the coasting photon thick
outflow is significantly wider that the thermal spectrum, see Fig.
\ref{CoastWindSpectrum}. Low energy part is described by a power law with Band
indices respectively $\alpha=0.34$ and $\alpha=0$, for sharp and fuzzy
photosphere approximations.

After initial phase of evolution, namely rising of the low-energy part, spectra
do not evolve until observer detects emission from the innermost part of the
outflow. At that moment there is a transition to another phase characterized by
the fast decrease of both temperature and flux. Considering time-integrated
spectrum we find that as characteristic times of the first and third phases are
much less than that of the second one, the spectrum is close to the
instantaneous one described above.

\subsection{Comparison with previous works}

Now we compare the results with obtained by other methods.
\cite{2011ApJ...737...68B} considered the photospheric emission from infinitely
long wind both at acceleration and coasting phases and solved the corresponding
\emph{steady} radiative transfer equation. His main conclusion is that in
addition to usual relativistic beaming leading to anisotropy of radiation in
laboratory frame, in the coasting wind another anisotropy in the comoving frame
of the outflow is developing. This comoving anisotropy results from the
fraction of photons which already underwent their last scattering in the bulk
photon field of the outflow. The anisotropy of such photons grows with
increasing radius for geometrical reasons. Since the amount of such photons
increase with radius the entire photon field becomes increasingly anisotropic.

For the finite photon thick outflow the radiative transfer problem becomes
explicitly time dependent. The expanding outflow at a given laboratory time
spans only a finite part of the probability density distribution shown in
Figs.\ref{AccelWindAppPhotosphere} and \ref{WindAppPhotosphere}, that results
in difference in observed spectra for finite and infinite cases. Interesting
consequence of ultrarelativistic motion of the outflow is that even
geometrically thin outflow $l\ll R$ at a given arrival time spans large
interval of laboratory radius $\Delta r=2\Gamma^2 l$.

The effect of additional comoving anisotropy on the source function found by
\cite{2011ApJ...737...68B} is actually small. The difference between the
probability of last scattering (\ref{PDF}) and the distribution of last
scattering in a steady wind does not exceed several percent.

Our method is also similar to the one used by \cite{2008ApJ...682..463P} and by
\cite{2011ApJ...732...49P} who described the late-time photospheric emission of
switching off relativistic wind considering the probability density function
(\ref{PDF}) for the last scattering of photons. An additional approximation
adopted by these authors is the possibility to split radial and angular
dependencies. Actually \cite{2008ApJ...682..463P} computes not the traditional
energy flux understood as energy crossing unit area in unit time, but photon
flux as number of photons crossing unit area in unit time. For this reason his
decay law for photon flux at late times is $F^{ob}(t_{a})\propto t_{a}^{-2}$.
Lorentz transformation of the photon energy from the comoving frame to the
laboratory one results in additional multiplier $(1-\beta\mu)^{-1}$ in the
energy flux that leads to the observed flux $F\propto t_{a}^{-3}$, which agrees
with our result in Eq. (\ref{swithchoff}), see also \cite{2011ApJ...732...49P}.

We conclude that the fuzzy photosphere approximation in fact follows closely
methods of \cite{2011ApJ...732...49P} and \cite{2011ApJ...737...68B}. In fact
we obtained similar results for the probability of last scattering as more
sophisticated treatment of radiation transfer \citep{2011ApJ...737...68B}. The
sharp photosphere approximation provides good description of light curves
including their raising and decaying parts. The observed spectrum from
accelerating outflow is also well described in this approximation, while there
is some difference for the coasting case. The advantage of sharp photosphere
approximation for computing observed light curves and spectra is evident for
intrinsically variable and dynamic outflows.

\section{Photospheric emission from photon thin outflows}

Now we turn to photon thin outflows. In Section \ref{diffusionSec} we pointed out
that most of the radiation leaves the outflow not at its photospheric radius,
but earlier, before the diffusion radius. Given that opacity of the outflow is
still large, the emission escapes only from a very narrow region near the outer
boundary of the outflow. The probability density function (\ref{PDF}) is
strongly peaked there and the photospheric emission for a given arrival time
originates from this narrow region. Sharp photosphere approximation is thus
completely justified in this case.

\begin{figure}[ptb]
\centering
\includegraphics[width=\columnwidth]{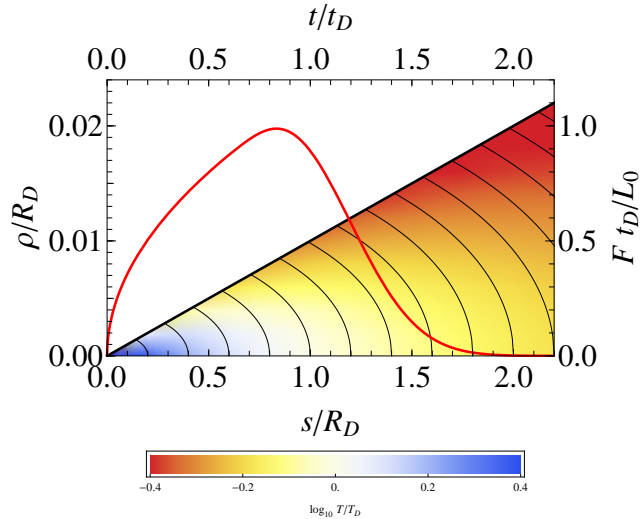}
\caption{Evolution of PhE for the photon thin coasting outflow and
dimensionless radiative diffusion flux, corrected for adiabatic cooling (red thick curve).
PhEs shown by thin curves correspond from left to right
to arrival times $t_a=(t_a^D/5, 2t_a^D/5, \dots)$,
see (\ref{tad}). Thick black curve bounding PhEs
correspond to the surface $\cos\theta=\beta$.
Relevant range of observed temperature of photospheric emission is
illustrated by color, see legend. Here $\Gamma=100$.
\label{CoastAppPhotosphere}}
\end{figure}

Since all the radiation is emitted from the PhE we briefly discuss its geometry
and dynamics. The PhE of the photon thin outflow is similar to EQTS of
infinitesimally thin constantly emitting relativistic shell considered firstly
by \cite{1939AnAp....2..271C} and then by \cite{1966Natur.211..468R,
1967MNRAS.135..345R}. The EQTS of this shell appears to a distant observer as
an ellipsoid with axes ratio equal to $\Gamma$. However the PhE of photon thin
outflow is not the entire ellipsoid: it is only a part of that surface, see
Fig. \ref{CoastAppPhotosphere}. The external boundary of the PhE for a given
$t_{a}$ is defined by the condition that photons emitted from the outermost
layer of the outflow toward observer leaves the outflow. In the photon thin
asymptotics this surface coincides with the relativistic beaming cone.

We again start with the radiative transfer equation (\ref{radtransf}). In
contrast with the photon thick case, here the source function $\mathcal I$ in
(\ref{sourcefunc}) strongly depends on both $r$ and $t$. The main process by
which photons are coupled to the matter is Compton scattering which conserves
the number of photons. Since opacity is large other processes which do not
conserve the photon number lead to local thermodynamic equilibrium with thermal
comoving radiation intensity $I_c$, number density and spectrum of photons in
the outflow. Hence we use the Rosseland radiative diffusion approximation (see
e.g. \cite{1979rpa..book.....R}, pp. 39--42), that we derive from the radiative
transfer equations (\ref{radtransf}) for expanding outflows. Details of
calculations are presented in Appendix.

It is useful to introduce the function $L_c(\xi,t)=(t/t_0)^{8/3} I_c(\xi,t)$
which accounts for the adiabatic cooling of radiation in expanding outflow.
Here both $\xi$ and time $t$ are measured in laboratory frame, while
$I_c(\xi,t)$ is measured in comoving frame. The diffusion equation with these
variables is well behaved in ultrarelativistic limit, see
\cite{2011ApJ...737...68B}. We then obtain
\begin{equation}%\label{}
\frac{\pd L_c}{\pd ct}-\frac{c^2t^2\Delta}{3R_0}\frac{\pd^2
L_c}{\pd\xi^2}=0, \qquad \Delta=\frac1{\Gamma^2\tau_0}.
\end{equation}
Notice that the diffusion coefficient is explicitly time dependent due to the
expansion of the outflow. In Appendix we also discuss initial and boundary
conditions and obtain an approximate analytic solution for the radiation field
inside the photon thin outflow, which is represented in Fig.
\ref{CoastAppPhotosphere}. The raising part of the corresponding flux of $L_c$
through the external boundary of the outflow scales as $t^{1/2}$, while its
decaying part is quasi-exponential one. Consequently, while the diffusion in a
static object gives the flux decreases as $t^{-1/2}$, in our case the observed
flux (\ref{Fnu}) is a more slowly decreasing function
\begin{equation}\label{phthinflux}
F\propto t_a^{-1/6},
\end{equation}
up to arrival time of diffusion
\begin{equation}\label{tad}
    t_a^D=\frac{R_D}{2\eta^2c}
        \simeq 0.12 E_{54}^{1/3}\eta_{2}^{-5/3} l_8^{1/3}\text{ s},
\end{equation}
where large part of energy has left the outflow already. At this moment the
energy decrease due to diffusion becomes substantial even in the deepest parts
of the outflow and later the observed flux decreases quasi-exponentially with
arrival time.

\begin{figure}[ptb]%
\centering
\includegraphics[
width=\columnwidth
]%
{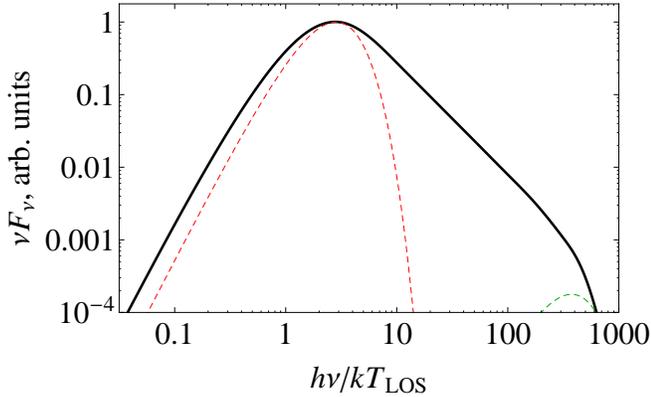}%
\caption{Time-integrated spectrum of photospheric emission of
photon thin outflow (thick curve, $\eta=100$, $R_{D}=10^5 R_0$),
superimposed with two instantaneous spectra of that emission,
corresponding to arrival time of photons emitted at the moment of
transition from acceleration to coasting (dashed green curve on the right)
and to arrival time of diffusion (dashed red curve on the left).
\label{CoastShellTISpectrum}}
\end{figure}

The comoving temperature of radiation on the photosphere is determined by the
balance between the energy diffusion from the interior of the outflow and
radiative losses and it is much smaller than the temperature in the interior.
The variation of observed temperature across the PhE is small, see Fig.
\ref{CoastAppPhotosphere} and hence the observed instantaneous spectrum is very
close to the thermal one and peaks near the observed temperature on the line of
sight. We find that the latter decreases as $t_a^{-13/24}$, in contrast with
adiabatic law $t_a^{-2/3}$. However, at diffusion radius both temperatures
coincide giving for the line of sight temperature
\begin{equation}\label{Tdiff}
    T_{LOS}\simeq 162 \eta_2^{4/9} \text{ keV}.
\end{equation}
The time integrated spectrum has a Band shape with a cut-off near the
temperature of transition from acceleration to coasting, see Fig.
\ref{CoastShellTISpectrum}. Low energy part of the spectrum has the slope
$\alpha=1$, while high-energy part has $\beta\simeq-3.5$.

\section{Discussion}

The expression (\ref{tad}) gives an estimate for duration of photospheric
emission of photon thin outflows. When available observed spectra are
integrated on time intervals comparable to (\ref{tad}) the observed spectrum of
photospheric emission is expected to have Band shape. Thus, starting from
comoving thermal spectrum for the photospheric emission we obtain for the first
time an observed spectrum which may be well described by the Band function with
high energy power law index $\beta$ being determined by the density profile of
the outflow. We find this result quite remarkable.

Notice that non thermal spectra as a result of convolution of thermal ones over
time has been introduced for afterglows of GRBs by \cite{1999ARep...43..739B}.
Double convolution over EQTS and arrival time is also one of the key ideas in
the fireshell model \citep{2003AIPC..668...16R}.

Band spectra in photospheric models of GRBs have been obtained by now only
assuming additional dissipation mechanisms such as magnetic reconnection
\citep{2006A&A...457..763G}, collisional heating \citep{2010MNRAS.407.1033B}
and internal shocks \citep{2011MNRAS.415.1663T,2011MNRAS.415.3693R}. In our
model such additional assumption is not required.

It is even more remarkable that GRBs appear to be the only known objects in nature
able to reach the photon thin asymptotics in their ultrarelativistic expansion.
For thermally accelerated relativistic plasmas which are discussed in
connection with their possible synthesis in ground based laboratories (e.g.
\citet{Ruffini2009}) it is unreachable. The photon thin asymptotics is reached
if the optical depth (\ref{tau0})
\begin{equation}
\tau_{0}\gg4\Gamma^{4}\frac{l}{R_0}=4\times10^{8}\eta_{2}^{4}\frac{l_8}{R_8}.
\end{equation}
GRBs clearly can satisfy this constraint as the contribution of baryons
$\tau_{0}\simeq3.5\times 10^{13}E_{54}\eta_{2}^{-1}R_{8}^{-1}l_{8}^{-1}$.

We obtained both time integrated and instantaneous observed spectra of
accelerating photon thick outflow which are close to thermal one with small
deviations in the Rayleigh-Jeans part, in agreement with
\cite{1998MNRAS.300.1158G}.

Time integrated observed spectrum of coasting photon thick outflow is broader
than thermal one, and deviates from it both in low- and high-energy parts. This
broadening is also found by \cite{2010MNRAS.407.1033B} using Monte-Carlo
simulations, but our spectral index of the low energy part is $\alpha=0$
instead of his $\alpha\simeq 0.4$. This difference may be attributed to the
fact that Monte-Carlo simulations do not account for stimulated emission.
Interestingly, we found $\alpha\simeq 0.34$ in the sharp photosphere
approximation.

As discussed earlier in Section \ref{optdepth} each photon thick outflow always
contains a photon thin layer with depth $\xi_{thin}=(R+l)/(2\eta^2)$ located
near its outer boundary. Radiation diffused out from this part of the outflow
arrives to observer first and modifies the initial part of the light curve and
the corresponding spectrum of the outflow. The diffusion length $\xi_{D}=
\left[(R+l)^3/(\eta^2\tau_0 R_0)\right]^{1/2}$ remains always within this
photon thin layer $\xi_{thin}$ and our solution for photon thin outflow is
applicable for description of this early emission. The photospheric emission
from well developed photon thick outflow and its late time behavior occurring
when the outflow crossed the photospheric radius may be described either by
fuzzy or by sharp photosphere approximations.

When the outflow becomes transparent in the transition from photon thick to
photon thin conditions, the observed time integrated spectrum will contain both
Band component produced by the early emission from the photon thin layer and
thermal-like component coming from the photon thick part superimposed. This may
be the reason why in most GRBs analyzed by \cite{2009ApJ...702.1211R} there are
both power law and black body components.

We presented analytic expressions for the photon flux in the simple model of
the portion of wind. With more complex density profile composed of presumably
many shells the light curve is expected to be variable and arbitrarily complex.
The minimal time scale of variability is given Eq. (\ref{tad}) and it may be
very small for small baryon loading. It is necessary to emphasize that the
decaying part of the light curve follows $t_a^{-3}$ for photon thick outflows.
Steeper decay of the light curve of photospheric emission it is a clear
signature of the photon thin outflow.

The photospheric emission should be additionally identified by the spectral
analysis. In particular, power law spectra extending to high energies above
10~MeV cannot be produced by the photospheric emission unless additional
mechanisms are involved. What we have shown here, though, is that the observed
spectrum may not necessarily be close to the thermal one.

\section{Conclusions}

In summary, in this paper we proposed a unified treatment of ultrarelativistic
outflows, which originate both from instantaneous and from continuous energy
release, with respect to photospheric emission. We have to stress that these
two cases are discussed in the literature and are respectively referred to as
shells and winds. Instead of this traditional division we propose a new
physically motivated classification, which in our opinion helps to understand
in particular why geometrically thin shell may appear as thick wind with
respect to the photospheric emission. For this reason we re-examined the
existing scattered literature and pointed out the advantage of the proposed
classification.

We studied geometry of photospheres in generic relativistic outflows. As we are
interested in appearance of the photosphere to a distant observer, we
introduced the notion of photospheric equitemporal surface and described its
dynamics.

We computed both energy flux and observed spectra of photon thick outflows in
two approximations, derived from the radiative transfer equation. In our fuzzy
photosphere approximation the effect of simultaneous emission from entire volume
of the outflow is taken into account. We also used another
computationally more simple sharp photosphere approximation which is shown to
reproduces well both light curves and spectra. These results generalize the
corresponding results in the literature for steady relativistic winds.

In photon thin outflows most of radiation is shown to originate
not at its photospheric radius, but at smaller radii due to radiation
diffusion. Starting from the radiative transfer equation for time dependent
outflows we derived the diffusion equation and obtained approximate analytic
solution for the energy flux. We present both instantaneous and time integrated
observed spectra. The latter are well described by the Band function. For our
simple density profile we find values for the low energy power law index
$\alpha=1$ and the high-energy power law index $\beta\simeq -3.5$.

\section*{Acknowledgment}

We thank the anonymous referee for useful remarks and suggestions.

%\bibliographystyle{apalike2}
%\bibliography{tapp}

\section*{Appendix}

For large opacities the distribution function of photons in comoving reference
frame is close to isotropic one and the radiative diffusion approximation is
accurate. Following \cite{2011ApJ...737...68B} we use spectral intensity in
comoving frame $J_\nu(t,\xi,\mu)$. Starting from the radiative transfer
equation (\ref{radtransf}) along the ray $s$ in laboratory frame, we transform
all variables except time $t$, depth $\xi$, and distance $s$ into comoving
reference frame
\begin{gather}%\label{}
    \nu^3\frac{d}{ds}\left(\frac{J_\nu}{\nu^3}\right)=
        \frac{\kappa_\nu}{\mathcal D}(S_\nu-J_\nu),
\end{gather}
and integrating over comoving frequency $\nu$ we have
\begin{gather}\label{transpeq}
    \frac1c\frac{\pd J}{\pd t}-\frac{\mu}{\Gamma \mathcal D}\frac{\pd J}{\pd \xi}
        +\frac{1-\mu^2}{vt-\xi}\frac{\pd J}{\pd \mu}+4\frac{\Gamma\beta}{\mathcal D}
        \frac{1-\mu^2}{vt-\xi}J=\frac{\kappa}{\mathcal D}(S-J),
\end{gather}
where $\mu=\cos\theta_c$, $\theta_c$ is the photon angle with respect to the
radial direction in comoving frame, $\mathcal D=\Gamma(1+\beta\mu)$ is Doppler
factor, $J=\int J_\nu d\nu$ is total photon intensity, $S=\int S_\nu d\nu$ is
total source function, $\kappa=J^{-1}\int\kappa_\nu J_\nu d\nu$ is effective
opacity, $\kappa_\nu=\sigma n_c$ is opacity in comoving frame.

In the case of small deviations from isotropy decomposition
\begin{equation}%\label{}
    J=J_0(t,\xi)+\mu J_1(t,\xi)
\end{equation}
could be applied. Introducing it into (\ref{transpeq}) and integrating it over
$\mathcal D d\mu$ and over $\mathcal D \mu d\mu$ after some algebra for
coherent scattering with $S=S_0=J_0$ we have
\begin{gather}%\label{}
    \label{J0}
    \frac{\pd J_0}{\pd ct}+\frac{\beta}{3}\frac{\pd J_1}{\pd t}
        -\frac{1}{3\Gamma^2}\frac{\pd J_1}{\pd \xi}
        +\frac{2J_1}{3(vt-\xi)}+\frac{4J_0\beta}{3(vt-\xi)}=0,\\
    \label{J1}
    \frac{\pd J_1}{\pd ct}+\beta\frac{\pd J_0}{\pd t}
        -\frac{1}{\Gamma^2}\frac{\pd J_0}{\pd\xi} +\frac{8J_1\beta}{5(vt-\xi)}
        =-\frac{\kappa J_1}{\Gamma}.
\end{gather}

Diffusion approximation is based on slow variation of total flux through the
entire sphere $L_1=J_1(t/t_0)^2$ over mean free path, so that $\frac{\pd
L_1}{\pd ct}=0$, and it provide $J_1$ from the equation (\ref{J1}). Inserting
this into (\ref{J0}) after simple but tedious calculations in ultrarelativistic
$\beta\simeq1$ photon thin case $\Gamma^2\xi\ll vt$ for function
$L=J_0(t/t_0)^{8/3}$ we obtain the diffusion equation
\begin{equation}%\label{}
    \frac{\pd L}{\pd ct}-\frac{c^2t^2\Delta}{3R_0}\frac{\pd^2 L}{\pd\xi^2}=0,\qquad
    \Delta=\frac{1}{\Gamma^2\tau_0}.
\end{equation}

This equation should be supplemented with boundary conditions. There are two
types of boundary conditions used frequently: free-streaming, for example in
two-stream approximation (\cite{1979rpa..book.....R}, pp. 42--45), and zero
boundary conditions, that can be used as replacement for free-streaming for
"extrapolated boundary" \citep{1994JOSAA..11.2727H}. We find that the position
of "extrapolated boundary" $\xi= -k\frac{c^2t^2\Delta}{R_0}$ ($k$ is a constant
of order unity, dependent on the approximation used for free-streaming
description) for the main part of emission is very close to the real boundary,
and in the case of zero boundary conditions $L|_{\xi=0}=L|_{\xi=l}=0$ there is
a series expansion of solution, that for initial conditions $L(\xi,t_0)=1$
gives
\begin{equation}\label{zeroboundsol}
    L(\xi,t)=\sum_{n=0}^{\infty}\frac{4}{(2n+1)\pi}
        \exp\left[-\frac{\Delta (2n+1)^2 \pi^2 c^3(t^3-t_0^3)}{9R_0l^2}\right]
        \sin\left[\frac{(2n+1)\pi\xi}{l}\right].
\end{equation}
This solution in comparison with numerical one with free-streaming boundary
conditions is accurate to a few percents.

The flux of $L$ is characterized by an initial burst and then tends to
the asymptotic solution, that corresponds to $t_0=0$, with flux
\begin{equation}\label{asympsol}
    F(t)=\frac{4\Delta c^3t^2}{3R_0l^2}\, \,
        \vartheta_2\left[0,
            \exp\left(-\frac{4\Delta \pi^2 c^3 t^3}{9R_0l^2}\right)
        \right],
\end{equation}
where $\vartheta_2$ is the Jacobi elliptic theta function, see Fig.
\ref{CoastAppPhotosphere}. The peak of the flux of $L$ is near the diffusion
time
\begin{equation}\label{td}
    t_D=\frac{l}{c}\left(\frac{R_0}{l \Delta}\right)^{1/3},
\end{equation}
and "extrapolated boundary" $\xi=-k l (l\Delta/R_0)^{1/3}\ll l$ is very close
to real one as $\Delta\ll1$, that ensures the accuracy of (\ref{zeroboundsol}).

\end{document}